\documentclass[final,1p,times]{elsarticle} % do not change
\usepackage{graphicx} % do not change
\usepackage{amssymb} % do not change
\usepackage{amsthm} % do not change
\usepackage{lineno} % do not change

\newcommand{\be}{\begin{equation}}
\newcommand{\ee}{\end{equation}}
\newcommand{\ba}{\begin{eqnarray}}
\newcommand{\ea}{\end{eqnarray}}
\newcommand{\bi}{\begin{itemize}}
\newcommand{\ei}{\end{itemize}}

\newcommand{\<}{\langle} 
\renewcommand{\>}{\rangle}  
\newcommand{\eq}{Eq.~}

\newcommand{\fig}{Fig.~}

\newcommand{\la}{\label}

\newcommand{\half}{{\textstyle\frac{1}{2}}}

\newcommand{\txts}{\textstyle}

\newcommand{\as}{a_{\sigma}}
\newcommand{\at}{a_{\tau}}
\newcommand{\Nt}{N_{\tau}}
\newcommand{\Ns}{N_{\sigma}}

\journal{Nuclear Physics A} % do not change
\begin{document} % do not change

\begin{frontmatter} % do not change

%% QM09Author: please enter your  
%% Title, author and address info here; please do not use footnotes

% Your Title - please insert
\title{Transport properties of the quark-gluon plasma from lattice QCD}

% Principle author, and co-authors - please insert
\author{Harvey~B.~Meyer}

% Address - please insert
\address{Center for Theoretical Physics \\
Massachusetts Institute of Technology\\
77 Massachusetts Ave \\
Cambridge, MA 02139, U.S.A.}

\begin{abstract} % do not change
I review the progress made in extracting transport properties of 
the quark-gluon plasma from lattice QCD simulations.
The information on shear and bulk viscosity,
the ``low-energy constants'' of hydrodynamics, is encoded 
in the retarded correlators of $T_{\mu\nu}$, the energy-momentum tensor.
Euclidean correlators, computable on the lattice, are related to 
the retarded correlators by an integral transform.
The most promising strategy to extract shear and bulk viscosity
is to study the shear and sound channel correlators
where the hydrodynamic modes dominate.
I present preliminary results from a comprehensive study 
of the gluonic plasma between $0.95T_c$ and $4.0T_c$.
\end{abstract} % do not change

\end{frontmatter} % do not change

%% QM09: we keep linenumbers at least for initial version
% \linenumbers % do not change

%% start of main text - please insert. 

%%\section{}\label{}

%%%%%%%%%%%%%%%%%%%%%%%%%%%%%%%%%%%%%%%
\section{Introduction}\label{intro}
%%%%%%%%%%%%%%%%%%%%%%%%%%%%%%%%%%%%%%%
The phenomenology of heavy ion collisions 
at the Relativistic Heavy Ion Collider (RHIC)
 has revealed unexpected properties of the quark gluon plasma.
Hydrodynamics calculations~\cite{Kolb:2000fha}
successfully described the distribution of 
produced particles in these collisions~\cite{Arsene:2004fa}.
This early agreement between ideal hydrodynamics
and experiment has been refined in recent times.
The dissipative effects of shear viscosity $\eta$
have been included  in (2+1)D hydrodynamics
calculations~\cite{Romatschke:2007mq}
% \cite{Romatschke:2007mq,Dusling:2007gi,Song:2007fn}
and the sensitivity to initial conditions quantitatively 
estimated~\cite{Luzum:2008cw} for the first time.
Experimentally, the elliptic flow observable $v_2$,
which is sensitive to the value of $\eta$ in units
 of entropy density $s$,
is now corrected for non-medium-generated two-particle 
correlations~\cite{:2008ed}.
The conclusion that $\eta/s$ must be much smaller than unity
has so far withstood these refinements 
of heavy-ion phenomenology~\cite{Song:2008hj}.
On the theory side,
it is therefore important to compute the QCD shear viscosity from first principles
in order to complete the picture.
Furthermore, since the heavy ion collision program at LHC will probe the quark-gluon
plasma at temperatures about a factor two higher~\cite{wiedemann-qm09}, 
it is crucial to predict
the shear viscosity at $\sim3T_c$, and to relate it to the size of elliptic flow,
before experimental data is available.

In this talk I present  lattice calculations of the
thermal correlators of the energy-momentum tensor (EMT)
in the Euclidean SU(3) pure gauge theory (i.e. quarkless QCD), and
discuss methods to extract the shear and bulk viscosity from them.
Computationally, the calculation is challenging enough 
without the inclusion of dynamical quarks,
and physically, the static properties of the QGP, normalized by the 
number of degrees of freedom, 
do not depend sensitively on the flavor content 
away from $T_c$~\cite{Karsch:2006sf}.
In perturbation theory~\cite{Arnold:2003zc}, the ratio of shear viscosity to
entropy density is O($\frac{1}{\alpha_s^2})$ and
there is only about $30\%$ difference between the pure gauge theory
and full QCD~\cite{Moore:2004kp} at a fixed value of $\alpha_s$
($\eta/s$ is smaller in the pure gauge theory). % , see \fig\ref{fig:moore}).
The bulk viscosity $\zeta$ is O($\alpha_s^2$) and 
predicted to be much smaller than the shear viscosity
in perturbation theory.
As a rule of thumb, $\eta/s\approx1.0$ \cite{Arnold:2003zc}
and $\zeta/\eta\approx10^{-3}$ \cite{Arnold:2006fz} at 
$\alpha_s=0.25$.

Since the pioneering calculation~\cite{Karsch:1986cq}, 
there have been only few attempts 
to calculate the viscosities of the pure gauge theory on the lattice.
Nakamura and Sakai~\cite{Nakamura:2004sy} performed the first calculations at 
$\Nt\equiv (aT)^{-1}= 8$, where $a$ denotes the lattice spacing. 
About two years ago, the accuracy
of the Euclidean correlators improved 
significantly~\cite{Meyer:2007ic,Meyer:2007dy} 
thanks to high statistics and a more efficient
two-level algorithm~\cite{Meyer:2002cd,Meyer:2003hy}.
Here I will give an update on the progress  made
in constraining dynamical properties of 
the QGP such as the viscosities.
I will not describe the calculations of 
the electric conductivity~\cite{Aarts:2007wj}
or heavy quark diffusion constant~\cite{Petreczky:2005nh}, except for saying that
they share many features with the viscosity studies.

%%%%%%%%%%%%%%%%%%%%%%%%%%%%%%%%
\section{Hydrodynamics and energy-momentum tensor correlators}
%%%%%%%%%%%%%%%%%%%%%%%%%%%%%%%%
From the modern point of view, hydrodynamics is an
effective theory that describes the slow, long-wavelength
motion of a fluid.
The central object is the energy momentum tensor,
a symmetric rank-two tensor, whose conservation is expressed
by the four continuity equations $\partial_\mu T^{\mu\nu}=0$.
The component $T_{00}$ is the energy density, $T_{0k}$
is the momentum density, which coincides with the energy flux,
and the spatial components $T_{jk}$ are the momentum fluxes.
From the microscopic point of view, % i.e. on the quantum field theory side, 
the matrix elements of $T_{\mu\nu}$, viewed as a quantum operator,
between any two on-shell states satisfy $\partial_\mu \<\Psi|T^{\mu\nu}(x)|\Phi\>=0$.

The two macroscopic modes of fluid motion near equilibrium
are the shear and sound modes.
% The simplest way to think about them is the following.
The former corresponds to the transverse diffusion
of momentum. Indeed
a small perturbation $T_{03} = T_{03}(t,x_1)$ of the
fluid around equilibrium 
gives rise to shear flow and satisfies the diffusion equation
\be 
 \partial_t \, T_{03}(t,x_1) - D \partial_1^2 \, T_{03}(t,x_1) =0\,,  
\qquad\quad 
\textrm{(shear mode)}
\ee
where the diffusion coefficient $D$ is 
proportional to the shear viscosity $\eta$, 
$D = \frac{\eta}{e+p}$ ($e$ is the energy density and 
$p$ the pressure).
Secondly, a sound wave propagating in the $z$-direction
with wavelength $\lambda = 2\pi/k$ is damped according to
\be  
T_{03}(t,k) \propto e^{-\frac{1}{2}(\frac{4}{3}\eta+\zeta)k^2t/(e+p)} 
\qquad\qquad \textrm{(sound mode)}\,.
\ee
Both the shear viscosity $\eta$ and bulk viscosity $\zeta$ 
thus contribute to the damping of sound waves.

The quantities computed on the lattice are correlators
of the energy-momentum tensor
that depend on Euclidean time $x_0$ 
and spatial momentum ${\bf q}$,
\be 
C_{\mu\nu,\rho\sigma}(x_0,{\bf q})
  =  L_0^5\int d^3{\bf x} ~e^{i{\bf q\cdot x}}
 ~\< T_{\mu\nu}(x_0,{\bf x}) T_{\rho\sigma}(0) \>\,,
 \qquad L_0\equiv 1/T\,.
\la{eq:Cx0}
\ee
The so-called spectral function
$\rho(\omega,{\bf q})$ is defined as ($-\pi\,\times$) the imaginary part 
of the Fourier-space, Minkowski-time retarded correlator $G_R(\omega,{\bf q})$.
I leave the temperature dependence of the functions $C$ and $\rho$ implicit.
For $\mu=\rho$ and $\nu=\sigma$,
it obeys a positivity condition, in our conventions 
$i^n\rho(\omega,{\bf p})/\omega\geq0$
where $n$ is the number of time components among the four indices,
and is odd in $\omega$, $\rho(-\omega,{\bf q})=-\rho(\omega,{\bf q})$.
Via a Kubo-Martin Schwinger relation,
this spectral function is related to the corresponding Euclidean correlator
(see~\cite{Petreczky:2005nh}),
\ba
C_{\mu\nu,\rho\sigma}(x_0,{\bf q}) &=& L_0^5
\int_0^\infty \rho_{\mu\nu,\rho\sigma}(\omega,{\bf q}) 
\frac{\cosh \omega(\half L_0-x_0)}{\sinh \half\omega L_0} d\omega\,.
\la{eq:C=int_rho}
\ea
A wealth of information is encoded in the spectral functions.
In particular, 
the shear and bulk viscosities are given by Kubo formulas
(see \cite{Teaney:2006nc} for a derivation),
\be 
\eta(T) = \pi \lim_{\omega\to0} \lim_{{\bf q}\to 0}
\frac{\rho_{13,13}(\omega,{\bf q})}{\omega},
\qquad
{\txts\frac{4}{3}}\zeta(T) +\eta(T)= \pi \lim_{\omega\to0}
\lim_{{\bf q}\to 0} \frac{\rho_{33,33}(\omega,{\bf q})}{\omega} \,.
\ee
They also determine the static structure of the 
plasma~\cite{Meyer:2008sn,Meyer:2008dt},
 \be 
 C_{\mu\nu,\rho\sigma}({\bf r},T) \equiv 
 \< T_{\mu\nu}(x_0,{\bf r})\, T_{\rho\sigma}(x_0,{\bf 0}) \> =
 \lim_{\epsilon\to0}\int \frac{d^3{\bf q}}{(2\pi)^3}\,
 e^{i{\bf q\cdot r}}
 \int_0^\infty d\omega \, e^{-\epsilon\omega}
 \frac{\rho_{\mu\nu,\rho\sigma}(\omega,{\bf q},T)}{\tanh\omega/2T}\,.
 \ee
In the following, I summarize the known analytic properties
of the spectral functions. The goal will be to exploit these 
properties to help us solve the integral equation~(\ref{eq:C=int_rho})
for the spectral function, given the numerically calculated 
Euclidean correlators.

%%%%%%%%%%%%%%%%%%%%%%%%%%%%%%%%%%%%%%%%%%%%%%%%%%%%%%%%%%%%%%%%%%%%%%%
% \subsection{Ward identities}
%%%%%%%%%%%%%%%%%%%%%%%%%%%%%%%%%%%%%%%%%%%%%%%%%%%%%%%%%%%%%%%%%%%%%%%
The  conservation of the EMT, $\partial_\mu T_{\mu\nu}=0$,
implies in particular, for ${\bf q}=q\hat e_3$,
\be
\omega^4\,\rho_{{00},{00}}(\omega,{\bf q})  =
-\omega^2\,{q}^2\,\rho_{{03},{03}}(\omega,{\bf q})
= q^4 \rho_{33,33}(\omega,{\bf q}),
\quad
-\omega^2\,\rho_{{01},{01}}(\omega,{\bf q}) = q^2\,\rho_{13,13}(\omega,{\bf q})\,.
\la{eq:0202e1}
\ee
The first set of spectral functions corresponds to the sound channel
and the second to the shear channel~\cite{Teaney:2006nc}.
I therefore  use the notation
$\rho_{\rm snd}(\omega,{\bf q})\equiv-\rho_{{03},{03}}(\omega,{\bf q})$
and $\rho_{\rm sh}(\omega,{\bf q})\equiv  -\rho_{{01},{01}}(\omega,{\bf q})$.

%%%%%%%%%%%%%%%%%%%%%%%%%%%%%%%%%%%%%%%%%%%%%%%%%%%
% \subsection{General classification of the EMT correlators\la{sec:classif}}
%%%%%%%%%%%%%%%%%%%%%%%%%%%%%%%%%%%%%%%%%%%%%%%%%%%
Kovtun and Starinets~\cite{Kovtun:2005ev} analyzed the general tensor structure
of EMT correlators. I briefly summarize 
their results.
For a generic relativistic quantum field theory in four dimensions,
there are five independent functions of $(\omega,{\bf q})$ 
that determine all 
thermal correlators of the EMT.
At $T=0$, this number is reduced to 
two\footnote{For a conformal field theory, the number of independent 
correlators is three at $T>0$, and just one at $T=0$.}~\cite{Pivovarov:1999mr}.
A physically motivated  choice of five independent functions is:
% \cite{Kovtun:2005ev}:
two functions corresponding to 
the two hydrodynamics modes, the shear mode ($G_1$ in the 
notation of~\cite{Kovtun:2005ev}) and sound mode ($2G_2+3C_L$);
one function associated with the $\<T_{12}T_{12}\>$ correlator
and momentum in the $z$ direction ($G_3$);
finally, the correlators $\<T_{\mu\mu}T_{\nu\nu}\>$ and 
$\<T_{\mu\mu}T_{00}\>$ (respectively proportional to 
$2C_T+C_L$ and $C_T+2C_L$)  determine the last two 
functions. As an application,
it is worth noting that  if $\theta_{\rm q}$ is the polar angle of ${\bf q}$,
\be 
-C_{03,03}(x_0,{\bf q})= L_0^5\int_0^\infty d\omega\left\{
 \rho_{\rm snd}(\omega,{\bf q})\,\cos^2\theta_{\rm q}+
 \rho_{\rm sh}(\omega,{\bf q})\,\sin^2\theta_{\rm q} \right\}
\frac{\cosh\omega (\half L_0-x_0)}{\sinh \half \omega L_0}.
\la{eq:tensor}
\ee
In the future this will allow us to exploit
also momenta that are not aligned along a lattice axis.

At $T=0$, $G_1=G_2=G_3$ and $C_T=C_L$, and they become 
functions of $(\omega^2-{\bf q}^2)$. 
As a corollary,
the extent to which these relations are obeyed at $T>0$ 
provides us with 
a way to evaluate the importance of thermal effects
on the correlation functions.

%%%%%%%%%%%%%%%%%%%%%%%%%%%%%%%%%%%%%%%%%%%%%%%%%%%%%%%%%%%%%%%%%%%%%%%
\begin{figure}
\centerline{\includegraphics[width=0.50\textwidth]{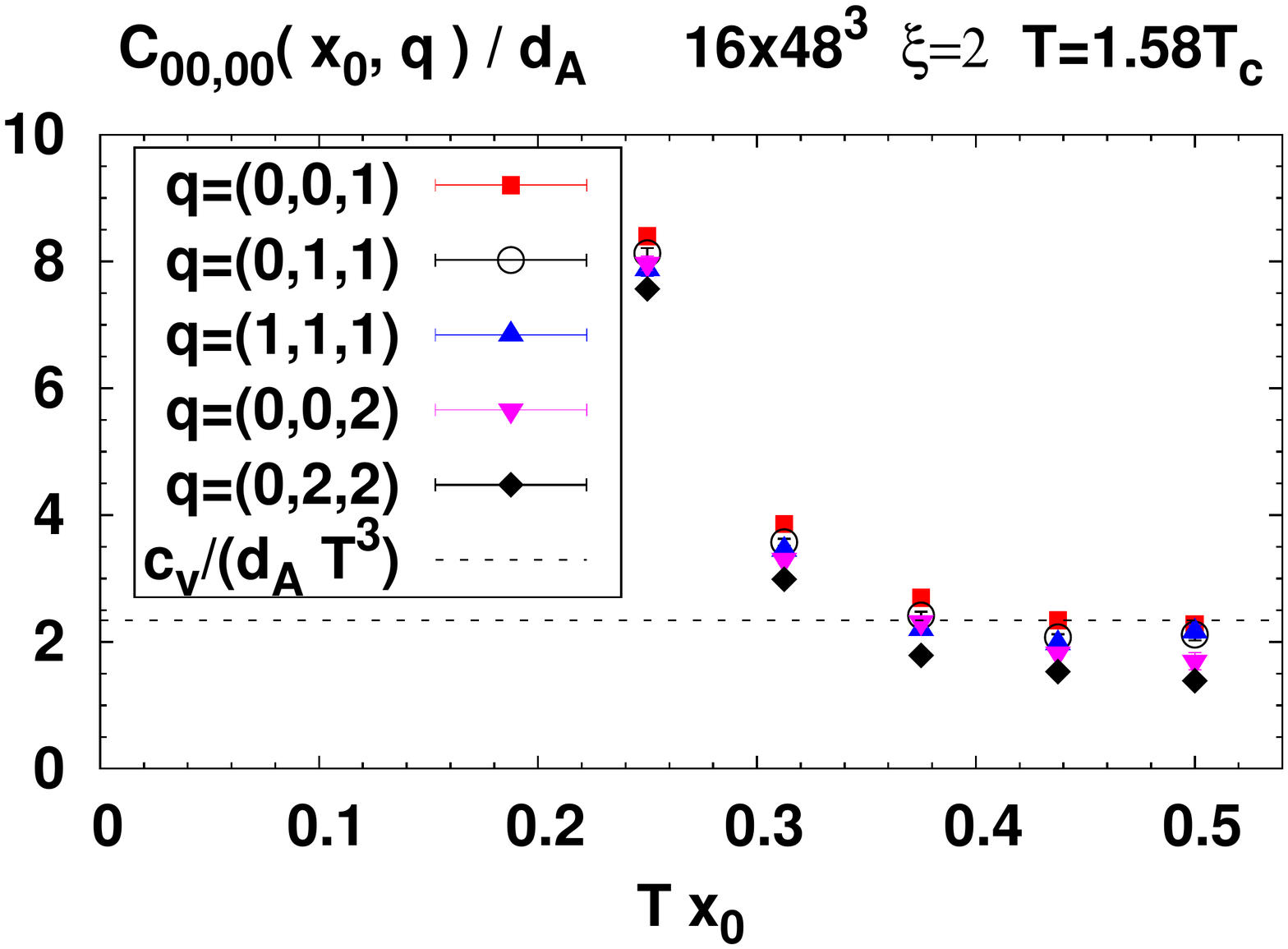}
\includegraphics[width=0.5\textwidth]{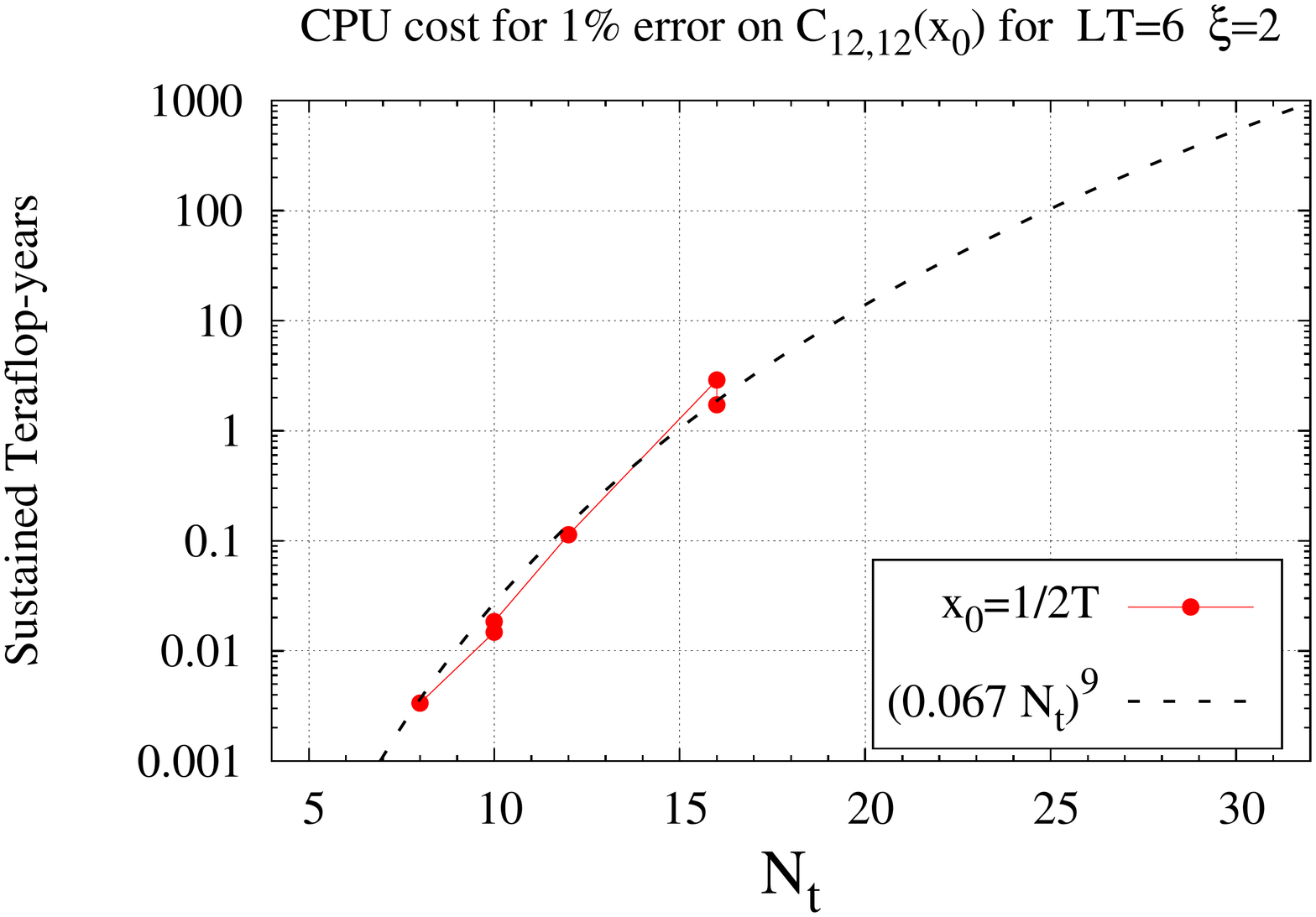}}
\vspace{-1.0cm}

\caption{
Left: Euclidean sound channel correlator.
Scaling of CPU-cost with $N_\tau$ in the deconfined phase.
}
\la{fig:Eucl}
\end{figure}
%%%%%%%%%%%%%%%%%%%%%%%%%%%%%%%%%%%%%%%%%%%%%%%%%%%%%%%%%%%%%%%%%%%%%%%
%%%%%%%%%%%%%%%%%%%%%%%%%%%%%%%%%%%%%%%%%%%%%%%%%%%%%%
% \subsection{The spectral functions in the low- and high-energy limits}
%%%%%%%%%%%%%%%%%%%%%%%%%%%%%%%%%%%%%%%%%%%%%%%%%%%%%%
As a low-energy theory, hydrodynamics predicts the small $k=(\omega,{\bf q})$
behavior of spectral functions in terms of a few `low-energy constants'.
% which are the transport coefficients.
A first-order hydrodynamic expression for the shear and sound 
spectral functions was derived in~\cite{Teaney:2006nc}.
For ${\bf q}=q\hat e_3$,
\ba
\frac{\rho_{\rm sh}(\omega,{\bf q})}{\omega} 
&\stackrel{\omega,q\to 0}{\sim} &
\frac{\eta}{\pi}\, \frac{q^2}{\omega^2+(\eta q^2/(e+P))^2}\,,
\la{eq:sf-hydro-shear}
\\
\frac{\rho_{\rm snd}(\omega,{\bf q})}{\omega} 
 &\stackrel{\omega,q\to 0}{\sim}&   
\frac{\Gamma_s}{\pi}\,
\frac{(e +P)\,q^2\omega^2}{(\omega^2-v_s^2q^2)^2+(\Gamma_s\omega q^2)^2}\,.
\la{eq:sf-hydro-snd}
\ea
I introduced the speed of sound $v_s$ and the 
sound attenuation length 
$\Gamma_s = (\frac{4}{3}\eta+\zeta)/(e+P)$.
Baier et al. \cite{Baier:2007ix} derived the dispersion relation
of the sound pole to next-to-leading order accuracy for a conformal theory:
\be
\omega = v_s(q) q,\qquad
v_s(q)=v_s \left\{1 + \frac{\Gamma_s}{2}q^2
\left(\tau_\Pi-\frac{\Gamma_s}{4v_s^2}\right)+{\rm O}(q^4)\right\}\,.
\la{eq:vsq}
\ee
Here $\tau_\Pi$ is the relaxation time for shear stress.

The Euclidean correlators and spectral functions 
were calculated to leading order at weak coupling 
in~\cite{Meyer:2008gt}. % meyerh
The correlators are also known to next-to-leading order 
at zero temperature~\cite{Kataev:1981gr,Pivovarov:1999mr}. % pivovarov
Expressing the latter results in terms of a shear channel spectral function,
\ba  
\rho_{13,13}(\omega,{\bf 0})
\stackrel{\omega\to\infty}{\sim} 
\frac{d_A\,\omega^4}{10(4\pi)^2}\left[1-\frac{5\alpha_s\,N_c}{9\pi}\right].
\ea

%%%%%%%%%%%%%%%%%%%%%%%%%%%%%%%%%%%%%%%%
% \subsection{Sum rules}
%%%%%%%%%%%%%%%%%%%%%%%%%%%%%%%%%%%%%%%%
Son and Romatschke~\cite{Romatschke:2009ng}
derived two sum rules for the spectral functions. 
The `bulk' sum rule can be reexpressed 
in terms of sound and shear channel spectral functions,
\be 
2\int_0^\infty \frac{d\omega}{\omega}\,
\Big[\rho_{33,33}(\omega,{\bf 0})-\rho_{13,13}(\omega,{\bf 0})\Big]_{T-0}
=(e+p)\Big(1-v_s^2-\frac{1}{9v_s^2}\Big)-\frac{4}{9}(e-3p)\,.
\ee
From a practical point of view, 
this provides an additional constraint on the spectral functions,
albeit at the price of having to  determine simultaneously the $T=0$ 
spectral functions. Because the symmetry group is larger
at $T=0$, % as detailed in  section~\ref{sec:classif}, 
the vacuum spectral functions can however be more 
strongly constrained by lattice data (in addition to having 
a longer Euclidean time extent).

%%%%%%%%%%%%%%%%%%%%%%%%%%%%%%%%%%%%%%%
% \subsection{Exact results at strong coupling from AdS/CFT\la{sec:adscft}}
%%%%%%%%%%%%%%%%%%%%%%%%%%%%%%%%%%%%%%%
In a class of strongly coupled gauge theories at large-$N_c$,
exact results have been obtained for the transport coefficients.
The most important result is that the shear viscosity 
assumes a universal value in a large class of theories,
$\eta/s=1/4\pi$~\cite{Kovtun:2004de}. The spectral functions for the 
EMT have also been obtained by numerical integration
in the case of the ${\cal N}=4$ SYM theory~\cite{Kovtun:2006pf}.
The result is compared to the hydrodynamic functional 
form~(\ref{eq:sf-hydro-snd})
in (\cite{Meyer:2008sn}, Fig. 3). Such a comparison teaches us up to what 
frequency and momentum the hydrodynamic spectral function
remains a good approximation to the exact result.
It turns out that up to $\omega,q\approx\pi T$, 
the sound channel spectral function is well described by hydrodynamics.
This sort of statement necessarily depends on 
the details of the microscopic theory; 
and in a weakly coupled theory, ones expects the validity
of hydrodynamics to be limited to much greater wavelengths.
Nevertheless I will use this observation as a guideline for the gluon plasma
at $T<4T_c$. A similar logic has also been applied in studies
of Mach cones in the plasma~\cite{chesler-QM09}.

%%%%%%%%%%%%%%%%%%%%%%%%%%%%%%%%%%%%%%%
\section{Lattice calculation}\label{latcal}
%%%%%%%%%%%%%%%%%%%%%%%%%%%%%%%%%%%%%%%
I use Monte-Carlo simulations of the anisotropic Wilson action 
for SU(3) gauge theory with a fixed anisotropy of $\xi=2$,
meaning that the temporal lattice spacing $\at$ is half the length
of the spatial lattice spacing $\as$.
See~\cite{Namekawa:2001ih} for details of the action.
The lattice spacing is related to the bare coupling 
(an input parameter of the simulation appearing in the lattice action) 
through $g_0^2\sim 1/\log(1/\as\Lambda)$.
Due to the loss of continuous translation invariance on the lattice,
the discretized EMT is not protected against 
a non-trivial $g_0$-dependent renormalization.
Fully calibrating the EMT on the anisotropic lattice
remains a numerical challenge. While $T_{0k}$ is normalized multiplicatively
and $T_{00}$ and $T_{\mu\mu}$ can be calibrated 
by using thermodynamic expectation values, 
the spatial components present the biggest challenge.

% \subsection{Numerical Euclidean correlators}

A typical example of a Euclidean correlator computed 
on the lattice is shown in \fig\ref{fig:Eucl}.
It displays the correlator of the energy density $T_{00}$,
at various non-vanishing spatial momenta. The lattice size 
is $16\times 48^3$ and the temperature about $2.3T_c$.
The typical size of the error bars is $3\%$.
Obtaining the correlators to that accuracy is 
computationally  demanding. Figure~\ref{fig:Eucl} also
shows how the cost of the computation scales with $\Nt$.
Different temperatures lie approximately on a smooth curve.
The dashed line corresponds to the expectation for a conformal
theory, $\tau_{\rm CPU}\propto \Nt^6\Ns^3$.
For a fixed aspect ratio $\Nt/\Ns$, the cost 
therefore scales as $\Nt^9$. This is in contrast with 
the expectation of an $\Nt^{11}\Ns^3$ scaling with
a one-level algorithm.

%%%%%%%%%%%%%%%%%%%%%%%%%%%%%%%%%%%%%%%%%%%%%%%%%%%%%%%%%%%%%%%%%%%%%%%
\begin{figure}
\centerline{
\includegraphics[width=0.35\textwidth]{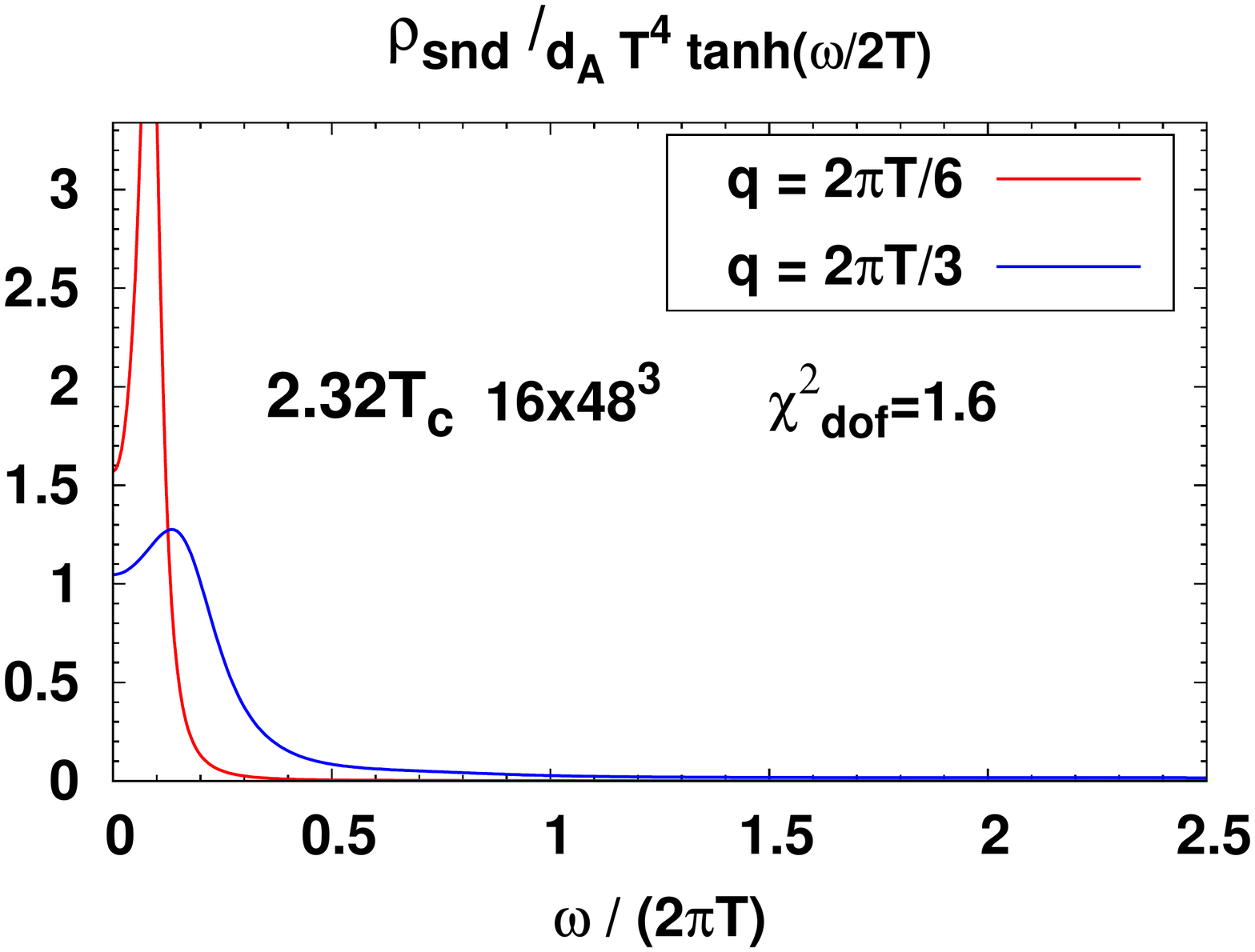}
\includegraphics[width=0.35\textwidth]{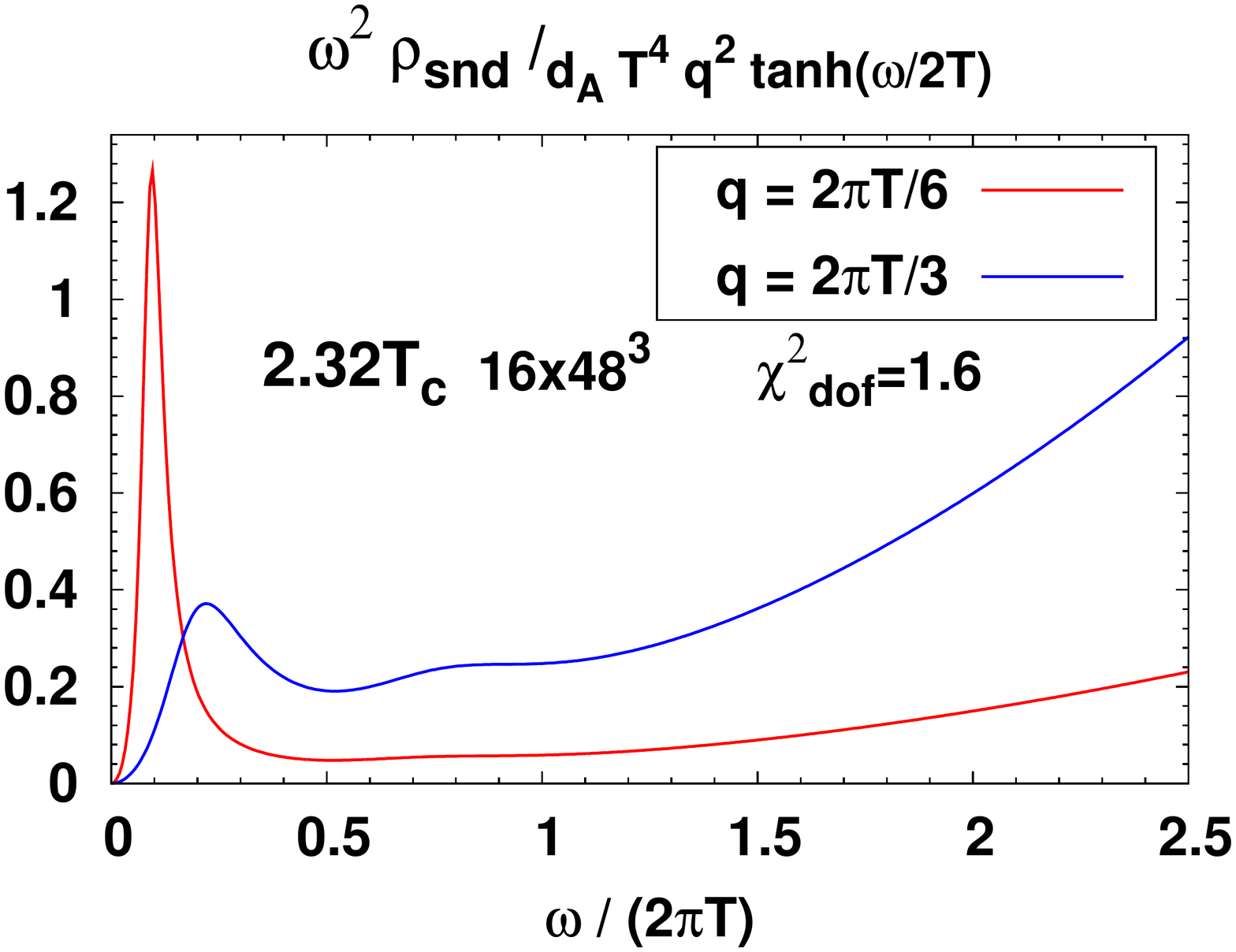}
\includegraphics[width=0.35\textwidth]{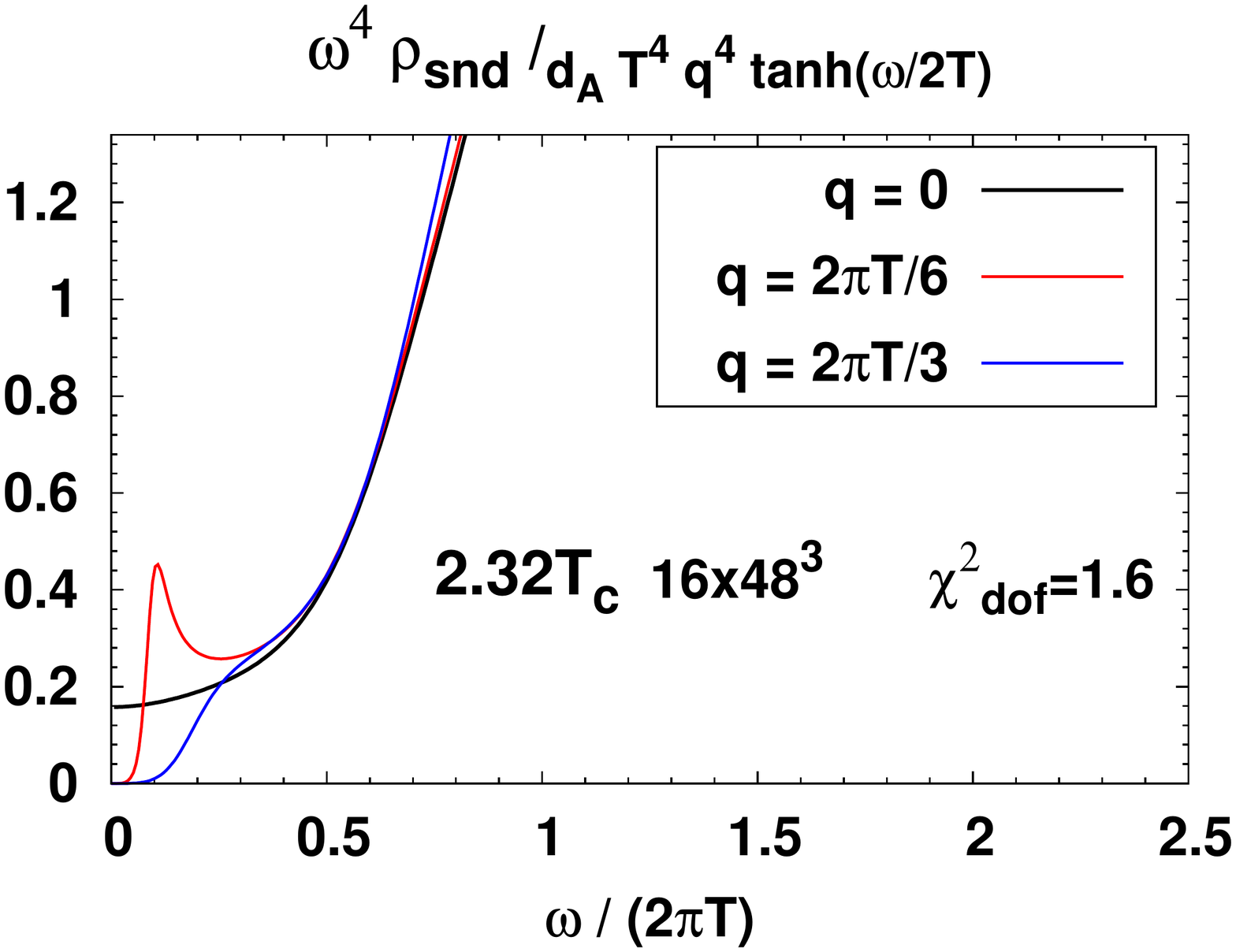}}
\centerline{
\begin{tabular}{c@{~~~~~}c@{~~~~~}c@{~~~~~}c@{~~~~~}c}
% \toprule
\hline
                     &    $1.58T_c$  &   $2.32T_c$ & free gluons & $\lambda=\infty$ SYM \\
                     \hline
$(\eta+\frac{3}{4}\zeta)/s$ & 0.20(3)  & 0.26(3) & $\infty$ & $\frac{1}{4\pi}\approx0.080$ \\
$2\pi T \tau_\Pi$    & 3.1(3)& 3.2(3) & $\infty $& $2-\log2\approx1.31$ \\
$(\eta+\frac{3}{4}\zeta)/(T\tau_\Pi s)$ & 0.40(5) & 0.51(5) &  0.17& 0.38 \\
% \bottomrule
\hline
\end{tabular}}
\caption{Sound channel spectral function  at $2.3T_c$. In the Table,
lattice results are compared to free gluons~\cite{York:2008rr} and to the strongly coupled
SYM results~\cite{Policastro:2001yc,Baier:2007ix}. Stat. errors only are given.
We expect $\zeta$ to be negligible at these temperatures.}
\la{fig:snd}
\end{figure}
%%%%%%%%%%%%%%%%%%%%%%%%%%%%%%%%%%%%%%%%%%%%%%%%%%%%%%%%%%%%%%%%%%%%%%%

The discretization errors affecting the Euclidean correlators 
calculated at finite lattice spacing were studied in 
perturbation theory in~\cite{Meyer:2009vj}. The study motivated 
in particular the choice of anisotropy $\as/\at\equiv\xi=2$.
Further it was shown that it is advantageous to remove 
the treelevel cutoff effects from the lattice correlators.
I apply this technique to all the data shown in these 
proceedings.
Finally, at treelevel the ${\bf q}=0$ correlator of $T_{0k}$
is much flatter as a function of $x_0$  than the correlator
of $T_{00}$ (they become exactly independent of $x_0$ in the 
continuum limit). This, and the fact that $T_{0k}$ 
is normalized multiplicatively even on the anisotropic lattice,
means that the correlators of the momentum density 
are the most accurately determined of all.

% \subsection{Test of the Ward identities}

Relations (\ref{eq:0202e1}) imply that in the sound channel,
the numerically determined 
correlators $C_{00,00}$, $C_{00,03}$, $C_{03,03}$,
$C_{33,03}$ and  $C_{33,33}$ 
are all related to the same spectral function $\rho_{\rm snd}$
modulo factors of $(\omega/q)$. Therefore a global
analysis of these five correlators is the best way to 
constrain $\rho_{\rm snd}$, provided the Euclidean correlators
are consistent with the Ward identities. This has to be the case
in the continuum limit, but presently analyses are still based 
on data at finite lattice spacing. It is therefore essential to 
check in at least some cases
that the Ward identities are satisfied to an accuracy
comparable to the statistical errors.
Conversely, these identities can also be used to 
determine some of the normalization factors of $T_{\mu\nu}$.
Figure (\ref{fig:nearTc}) compares the correlators
$C_{03,03}(x_0,{\bf q})$ and $C_{00,33}(x_0,{\bf q})$
for several values of $x_0$ and ${\bf q}=q\hat e_3$
(they are exactly equal in the continuum).
While the normalization of $T_{03}$ is easily determined by requiring
$C_{03,03}(x_0,{\bf 0})=s/T^3$, the operator $T_{33}$ 
requires in total four unknown normalization factors.
Two of them are fixed by thermodynamics, and two
have been fitted so that the two correlators
on \fig(\ref{fig:nearTc}) agree. Since there are 
O(20) points where the correlators are compared, this represents 
a non-trivial check.

%%%%%%%%%%%%%%%%%%%%%%%%%%%%%%%%%%%%%%%%%%%%%%%%%%%%%%%%%%%%%%%%%%%%%%%
\begin{figure}
\centerline{
\includegraphics[width=0.5\textwidth]{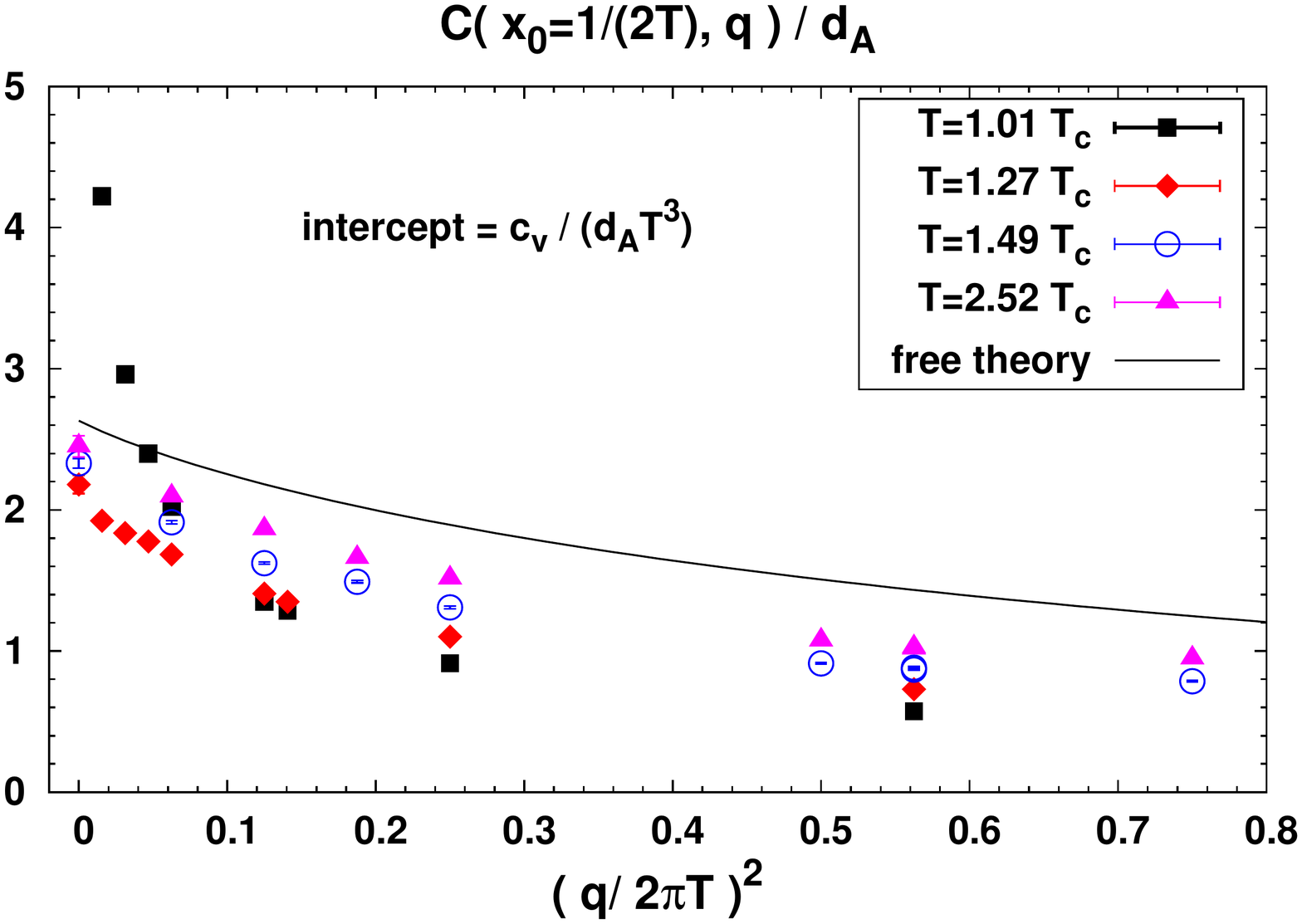}
\includegraphics[width=0.5\textwidth]{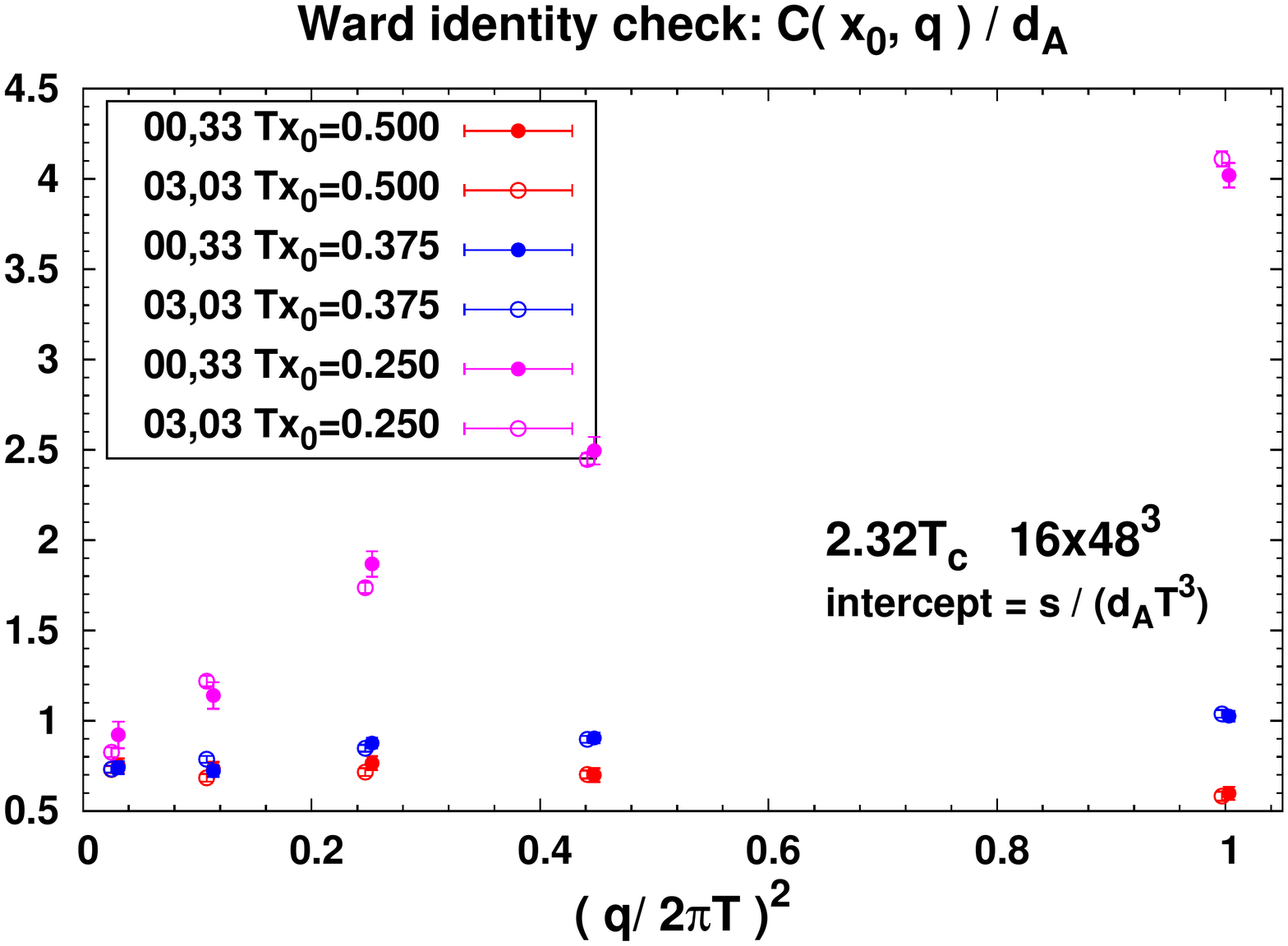}}
\vspace{-0.6cm}

\caption{
Left: $C_{00,00}(x_0,{\bf q})$ above and near $T_c$,
where evidence for the damping of sound waves is seen.
Right: test of a Ward identity at finite lattice spacing.}
\la{fig:nearTc}
\end{figure}
%%%%%%%%%%%%%%%%%%%%%%%%%%%%%%%%%%%%%%%%%%%%%%%%%%%%%%%%%%%%%%%%%%%%%%%
%%%%%%%%%%%%%%%%%%%%%%%%%%%%%%%%%%
% \subsection{A preliminary analysis of the sound channel}
%%%%%%%%%%%%%%%%%%%%%%%%%%%%%%%%%%
In the following I present an analysis of the sound channel correlators 
on a $16\times48^3$ lattice
for $x_0\geq 1/4T$ (corresponding to $x_0/\at\geq 4$)
and for $q\leq\pi T$. I assume for now, motivated by the inspection
of the strongly coupled ${\cal N}=4$ SYM spectral functions,
that hydrodynamics describes the low-frequency part of the 
spectral function up to that momentum.

I adopt a fit ansatz motivated by 
perturbation theory at high frequencies ($\rho_{\rm high}$), 
and hydrodynamics at low energies ($\rho_{\rm low}$). Specifically,
% \be 
 $\rho_{\rm snd} = \rho_{\rm low} + \rho_{\rm med} + \rho_{\rm high}$,
%   \ee
where
\ba
\frac{\rho_{\rm low}(\omega,q,T)}{\tanh(\omega/2T)} &=& 
\frac{2{\widehat\Gamma_s}}{\pi}
\frac{(e +P)\,\omega^2q^2}{(\omega^2-v_s^2(q)q^2)^2+({\widehat\Gamma_s}\omega q^2)^2}
\frac{1+{\sigma_1} \omega^2}{1+{\sigma_2} \omega^2}\,
\la{eq:rho-low} \\
\frac{\rho_{\rm med}(\omega,q,T)}{\tanh(\omega/2T)} &=& 
 \omega^2q^2\tanh^2(\frac{\omega}{2T}) 
\frac{{\ell \,\sigma}}{{\sigma}^2 + (\omega^2-q^2-{M}^2)^2}\,,
\la{eq:rho-med} \\
\frac{\rho_{\rm high}(\omega,q,T)}{\tanh(\omega/2T)} &=& 
\omega^2q^2\tanh^2(\frac{\omega}{2T}) \frac{2d_A}{15(4\pi)^2} \,,
\la{eq:rho-high}
\ea
where $v_s(q^2)$ is given by \eq\ref{eq:vsq}.
Some remarks are in order. Perturbation theory and the 
operator product expansion can be used to
systematically improve the knowledge of $\rho_{\rm high}$.
% For instance, the two-loop results of Pivovarov 
% et al.~\cite{Kataev:1981gr,Pivovarov:1999mr} can be used 
% to fix the leading term at high frequencies.
Secondly, a full second-order hydrodynamics 
parametrization of the spectral function at low-frequencies
would significantly improve our understanding of the accuracy
on the transport coefficients that can be expected from 
such as global fit.
Thirdly, the region $\omega={\rm O}(T)$ is the one where
there is least theoretical guidance. Given the present amount
and quality of the data, the simple ansatz 
(\ref{eq:rho-med}) used here
is a reasonable choice (other functional forms could be used to 
test the sensitivity of the transport coefficients 
to this particular choice). 
As the quality of the data improves,
the goal is to treat this part of 
the spectral function ($\rho_{\rm med}$)
in a more  systematic way,
 either by expanding it in a basis of orthogonal 
functions~\cite{Meyer:2007ic} or by using
 the Maximum Entropy Method~\cite{Asakawa:2000tr}.

In the present specific example 
I fit 7 parameters to a total of 48 data points.
The parameters are $\widehat\Gamma_s,\sigma_1,\sigma_2,
 \tau_\Pi, \ell, \sigma, M$.
Figure (\ref{fig:snd}) displays the reconstructed 
sound spectral function
based on lattice data at $2.3T_c$,
and the results for the fit parameters 
of interest are given there too.

%%%%%%%%%%%%%%%%%%%%%%%%%%%%%%%%%%%%%%%
% \section{Disappearance of the sound mode close to $T_c$}
%%%%%%%%%%%%%%%%%%%%%%%%%%%%%%%%%%%%%%%

There has been significant interest in the behavior of bulk
viscosity near the phase 
transition~\cite{Kharzeev:2007wb,Meyer:2008dq,Huebner:2008as}.
Figure (\ref{fig:nearTc}, left panel) displays the $C_{00,00}$
correlator as a function of ${\bf q}^2$.
Recalling the relation between the Euclidean correlator and 
the spectral function (\eq\ref{eq:C=int_rho}) as well as the positivity 
property of the latter, this Figure shows that for $T=1.01T_c$
and $q=\pi T/4$, $\frac{2}{d_AT^4}\int_0^{{\rm O}(T)} 
\frac{d\omega}{\omega}\rho_{\rm snd}(\omega,{\bf q})\ll 
c_v/d_AT^3\approx 13.6(1.9)$
($c_v$ is the specific heat). The sound peak
has therefore quasi disappeared 
even for a sound wavelength of 
$\lambda=8/T\approx 5.8$fm.
I conclude that for all practical purposes,
the gluonic medium does not support sound waves near $T_c$.

\section{Conclusion}
I have described the calculation of 
transport properties of the quark gluon plasma
on the lattice. For the time being the calculations 
are performed in the purely gluonic plasma
in order to reduce the computational cost.
The primary quantities computed are the Euclidean-time
dependent correlators of the energy-momentum tensor.
Based on these the spectral functions can be constrained,
particularly with the help of information on their functional form.
Hydrodynamics and perturbation theory
describe (in opposite regimes) both the spatial momentum
and frequency dependence of the correlators.
Correlators with finite spatial momentum 
are then useful, because within either the shear or the sound channel
they are interrelated 
by the energy and momentum conservation equations.
Different correlators can thus be used to over-constrain 
the shear and sound spectral functions.
Altogether, it is now possible 
to include 50-100 data points in a global analysis
of the sound channel, and, as illustrated by \eq\ref{eq:tensor},
even more points per channel can be used if 
the shear and sound channels are analyzed simultaneously.

My current best guess for the viscosity of the hot matter
that will be created at LHC uses the  lattice 
result $\eta/s\approx0.26$ for the purely 
gluonic plasma at 2.3$T_c$, and multiplies it by the 
perturbative (AMY) ratio of $\eta/s$ in full QCD and in 
the pure gauge theory~\cite{Arnold:2003zc}, %,see \fig\ref{fig:moore}),
\be 
 [\eta/s]_{QGP} \approx
[\eta/s]_{GP,lattice} 
\cdot \left[\frac{[\eta/s]_{QGP}}{[\eta/s]_{GP}}\right]_{\rm AMY}
\approx 0.40.\ee

I have presented direct evidence for the disappearance of 
sound waves as collective excitations of the plasma in the 
vicinity of the deconfining phase transition.
This effect leaves a strong signature on the Euclidean correlators
in the pure gauge theory, and it would be very interesting to know
how strong it remains in full QCD, where the transition to the 
QGP is a crossover. As the chiral critical point is approached
at finite baryon density
(assuming it exists), the suppression of sound waves becomes
stronger. This has recently been studied in detail 
by a combination of hydrodynamics
and the theory of critical phenomena~\cite{Minami:2009hn},
and even been proposed as a possible signature of the chiral critical point.

%\section*{Acknowledgments} % please insert, comment out or delete if not needed
I thank the organizers of the Quark Matter 2009 conference 
for a very enjoyable and stimulating physics event and for 
the opportunity to present this work.
Lattice computations were  carried out on facilities of
the USQCD Collaboration, which are funded by the Office of Science of
the U.S. Department of Energy, as well as on the 
Blue Gene L rack and the desktop machines of the
Laboratory for Nuclear Science at M.I.T. 
This work was supported in part by
funds provided by the U.S. Department of Energy 
under cooperative research agreement DE-FG02-94ER40818.

 % do not change 
\end{document}